\documentclass[traditabstract]{aa}
\usepackage[authoryear]{natbib}
\usepackage{graphicx}
\usepackage{amsfonts}
\def\lunits{$\rm erg\,s^{-1}$~}
\def\funits{$\rm erg\,cm^{-2}\,s^{-1}$~}
\def\cunits{$\rm cm^{-2}~$}

\def\chandra{{\it Chandra~}}

\begin{document}
\title{X-ray observations of sub-mm LABOCA galaxies in the eCDFS}


  \titlerunning{X-ray observations of sub-mm galaxies}
    \authorrunning{I. Georgantopoulos et al.}

   \author{I. Georgantopoulos\inst{1,2}
           E. Rovilos \inst{3}
           A. Comastri\inst{1} 
                     }

   \offprints{I. Georgantopoulos, \email{ioannis.georgantopoulos@oabo.inaf.it}}

   \institute{INAF-Osservatorio Astronomico di Bologna, Via Ranzani 1, 40127, Italy \\
              \and 
              Institute of Astronomy \& Astrophysics,
              National Observatory of Athens, 
 	      Palaia Penteli, 15236, Athens, Greece \\
              \and
              Max Planck Institut f\"{u}r extraterrestrische Physik,
              Giessenbachstra\ss e, 85748 Garching, Germany\\
             }

   \date{Received ; accepted }

\abstract{
We explore the X-ray properties of the 126 sub-mm galaxies (SMGs) of the LABOCA survey in the CDFS and the eCDFS regions. 
 SMGs are believed to experience  massive episodes of star-formation. Our goal is to examine whether star-formation 
  coexists with AGN activity,  determine the fraction of highly obscured AGN  and finally to obtain an idea of the dominant power-mechanism 
  in these sources.   
 Using Spitzer and radio arc-second positions for the SMGs, we find 14 sources with significant X-ray detections.  
  For most of these there are only photometric redshifts available, with their median redshift being $\sim2.3$. 
    Taking into account only the CDFS area which has the deepest X-ray observations, we estimate an X-ray AGN fraction of 
      $<26\pm 9$ \% among SMGs. 
 The X-ray spectral properties of the majority of the X-ray AGN which are associated with SMGs are 
  consistent with high obscuration, $>10^{23}$\,\cunits,
  but there is no unambiguous evidence for the presence of Compton-thick sources. 
   Detailed Spectral Energy Distribution fittings show that the bulk of total IR luminosity originates in star-forming processes,
    although a torus component is usually present.  
  Finally, stacking analysis of the  X-ray undetected SMGs  reveals a signal  in the soft (0.5-2 keV) and marginally in the hard (2-5 keV) X-ray band. 
   The  hardness ratio of the stacked signal is relatively soft ($-0.40\pm0.10$) corresponding to $\Gamma \sim1.6$. 
   This argues against a high fraction of Compton-thick sources among the X-ray undetected SMGs. 
     }
   \maketitle
%

\section{Introduction} 
The advent of the SCUBA detector (Holland et al. 1999) on the James Clerk Maxwell Telescope 
  has brought a spectacular advance in the 
 field of sub-mm cosmology. 
  The first extragalactic surveys at 850 $\rm \mu m$ revealed a population of numerous 
   luminous high redshift sub-millimeter galaxies or SMGs (see Blain et al. 2002 for a review). 
    Sub-mm surveys are very fruitful in detecting distant galaxies because of the negative 
    K-correction at sub-mm wavelengths  which counteracts the dimming of light with increasing distance.  
      The SMGs  generate significant fractions of the energy produced by all galaxies 
       over the history of the Universe (e.g.  Smail, Ivison \& Blain 1997, Barger et al. 1998, Hughes et al. 1998, Blain et al. 1999, 
        Arexatga et al. 2007, Wall, Pope \& Scott 2008), being probably among the most luminous objects in the Universe. 
         The limited spatial resolution at sub-mm wavelengths has hampered the identification of their optical counterparts. 
          Recently,  Chapman et al. (2003, 2005) have made great advances by  identifying a large number of SMGs through their 
           radio counterparts  and subsequently following them with Keck spectroscopy. 
        These studies have shown that the SMG population lies at high redshift with the median being z=2-3, 
         (see also Maiolino 2008 and references therein), 
         although the highest redshift SMGs have been found out to z$\sim4-5$ (Capak al. 2008, Daddi et al. 2009, Coppin et al. 2009, Wardlow 
          et al. 2010). 
                    
                       Mid-IR spectroscopy with Spitzer IRS  reveals mainly star-forming spectra 
            with implied star-formation rates as high as $1000\,\rm M_\odot\,yr^{-1}$ (Valiante et al. 2007, 
             Pope et al. 2008, Menendez-Delmestre et al. 2007, 2009).  
             Pope et al. (2008) obtained IRS 
              spectra of 13  SMGs in the CDFN:  11 of these were 
              detected with their spectra being dominated by PAH features. Given the known 
             anti-correlation   between PAH emission and AGN activity this directly suggests that star-formation 
                is the major powering mechanism in these SMGs. The median luminosity of these
                 powerful star-forming galaxies approaches $\rm 1\times 10^{13}\,L_\odot$\
                  at a median redshift of z=2.7. 
             
             The X-ray data have identified many AGN among SMGs.   
             Alexander et al. (2005a, 2005b) have found X-ray counterparts to a sample of 
             20 SMGs with radio counterparts from Chapman et al. (2005) in the Chandra Deep Field North (CDFN).
              Based on the number of X-ray detections, 
              Alexander et al. (2005b) claim a high fraction ($75\pm 19 \%$) of AGN 
               among the sub-mm galaxies which have radio counterparts. 
                If one takes into account the radio undetected SMGs in the CDFN,
                 making the conservative assumption that none of the radio undetected 
                  SMGs hosts AGN activity, then the AGN fraction in the SMG population 
                   becomes $>38^{+12}_{-10}$ \% (Alexander et al. 2005a).  
        Alexander et al. (2005b) claim that the vast majority of the radio-detected SMGs
          are highly obscured with 
        column densities exceeding $10^{23}$\,\cunits. The above results 
         suggest that  in SMGs the intense star-formation goes hand in hand 
          with the supermassive black-hole growth. 
           The low $\rm L_X/L_{FIR}$ ratio suggests that intense star-formation activity 
            dominates the bolometric output (Alexander et al. 2005b).   
            Laird et al. (2010) presented an analysis of 35 SMGs in the CDFN with sub-arcsec 
             positions either from radio or {\it Spitzer} counterparts. They find 16 objects with significant 
              X-ray detection, or a fraction of about $45\pm 8$ \%. However, they find that a dominant 
               AGN contribution is only required in only seven sources, or 20\% of the SMG sample.
       This figure lies at the lower limit of previous estimates raising questions 
        on how common the AGN contribution in sub-mm galaxies is.  
                 
                 Recently, the LABOCA sub-mm camera on the APEX telescope has performed 
            a sensitive sub-mm survey of the CDFS and its environs, that is the extended CDFS (eCDFS). 
            The CDFS is the region of the sky with the deepest X-ray data available 
             (together with the CDFN) and thus provides the opportunity 
              to further explore  the X-ray properties of SMGs.  
 Lutz et al. (2010) have performed a stacking analysis in the sub-mm LABOCA map 
  of the X-ray  sources detected in the eCDFS. Their main aim 
   is to explore the star-formation properties of the X-ray AGN population 
    as revealed by its sub-mm emission.
In this paper, in a complementary study we explore the X-ray 
 properties of   the SMGs in the LABOCA survey. Our goal,
  in analogy with the work of Alexander et al. (2005a) and Laird et al. (2010) 
   in the CDFN, is to study the fraction of  AGN among SMGs 
    and in particular the fraction of heavily obscured sources
    and Compton-thick AGN among them.  
    Hereafter,  we adopt $\rm H_o
=  75  \,  km  \,  s^{-1}  \,  Mpc^{-1}$,  $\rm \Omega_{M}  =  0.3$,\
$\Omega_\Lambda = 0.7$ throughout the paper.

\section{Data}

\subsection{The LABOCA sub-mm survey} 
 LABOCA is the sub-millimeter camera (Siringo et al. 2009) at the APEX telescope (G\"{u}sten et al. 2006). 
 We use the $\rm 870\,\mu m$  map obtained by the LABOCA eCDFS sub-mm survey  -LESS-  (Wei\ss et al. 2009). 
 LESS covers the full $30\times30$ arcmin field size of the eCDFS and has a uniform noise level of 
  $\rm \sigma_{870 \mu m} \approx 1.2 mJy/beam$. 
 This catalog contains 126 sources detected at $>3.7\sigma$.  Simulations show 
  that at the above extraction limit,  95\% of the sources have  positional accuracy  
  better than  8 arcsec (Wei\ss et al. 2009).
 
 \subsection{X-ray Data}
The 2Ms CDFS observations consist of 23 pointings. 
The analysis of the 1\,Ms data is presented 
 in Giacconi et al. (2002) and Alexander et al. (2003) while
  the analysis of all 23 observations is presented in 
 Luo et al. (2008).  
 The average aim point is $\alpha = 03^h 32^m
28^s.8$,  $\delta =  -27^\circ 48^{\prime}  23^{\prime\prime}$ (J2000).    
 The  23 observations cover an area of 435.6\,arcmin$^2$. 
 462 X-ray sources have been detected by Luo et al. (2008).  
 The CDFS survey reaches a sensitivity limit of $1.3\times 10^{-16}$\,\funits
and $1.9\times 10^{-17}$\,\funits in the hard (2-8\,keV) and soft (0.5-2\,keV) band
  respectively.  This corresponds to a false-positive probability 
  threshold of $10^{-6}$ (see Luo et al. 2008).
 The Galactic column density towards
the CDFS is $0.9\times 10^{20}$\,\cunits (Dickey \& Lockman 1990).

The eCDFS consists of four contiguous 250\,ks \chandra observations 
covering $\rm \approx 0.3\,deg^{2}$ centered on the CDFS. 
 Source detection has been performed by Lehmer et al. (2005) 
  and Virani et al. (2006). Here, we use the source detection 
   of Lehmer et al. (2005) in which 762 sources have been detected. 
    The sensitivity limit reaches $1.1\times 10^{-16}$ and 
     $6.6\times 10^{-16}$\,\funits in the 0.5-2 and 2-8\,keV 
      bands respectively.    This corresponds to a false-positive probability 
  threshold of $10^{-6}$. 

\subsection{Mid-IR} 
The central regions of the CDFS have been observed in
the mid-IR by the {\it Spitzer} mission (Werner 2000) as part of the Great
Observatory Origin Deep Survey (GOODS). These observations cover areas of
about $10 \times 16.5\rm \, arcmin^2$ in both fields using the IRAC (3.6, 4.5,
5.8 and  8.0$\, \rm \mu m$) and the  MIPS ($24\,  \rm  \mu m$) instruments onboard
{\it Spitzer}. The sensitivity is 
$\rm 80\,\mu{Jy}$ ($5\sigma$) in the $\rm 24\,\mu{ m}$ MIPS band. 
The data products are available from the {\it Spitzer} data centre
({http://data.spitzer.caltech.edu/popular/goods/}).

For the extended region bracketing the GOODS area we have used data from
\emph{Spitzer}'s IRAC and MUSYC Public Legacy of the eCDFS
(SIMPLE\footnote{{http://www.astro.yale.edu/dokkum/SIMPLE/}}, see
van Dokkum et al. 2005) and the Far-Infrared Deep Extragalactic Legacy survey
(FIDEL\footnote{{http://www.noao.edu/noao/fidel/}}, see
 Magnelli et al.  2009). SIMPLE is an IRAC survey of a $0.5\times 0.5$\,deg area
centered on the CDFS. The source catalogue contains more than 60000 sources with
an exposure time ranging from 2 to 4 hours and a flux limit of
$\sim24$\,mag(AB) in the 3.6\,$\mu$m band. FIDEL is a MIPS survey of the eCDFS
with a median exposure time of 8000\,s in the 24\,$\mu$m band and in the
$30\times 30$\,arcmin area of the eCDFS. 
We note that the area of the eCDFS has been also observed with the 
 {\it Herschel} far-IR mission (e.g. Shao et al. 2010) 
  but the data are not yet publicly available.  

\subsection{Optical Data}
  The CDFS has been imaged extensively in the optical 
   (e.g.   Giacconi et al. 2002). 
  In this paper we are making use of the CTIO-4\,m-MOSAIC\,II camera 
   observations  (Gawiser et al. 2006)
  taken as part of the MUSYC project.
 The survey is complete to a total magnitude of R=25(AB).

\section{Sample Selection}

A simple cross-correlation between the positions of the 126 LABOCA sources with
those of X-ray sources from the catalogues of  Luo   et al. (2008) and
 Lehmer et al. (2005) in the CDFS and eCDFS fields respectively, reveals 20
possible associations, using a search radius of 8\arcsec, the same as the
minimum positional accuracy of the LABOCA catalogue. In order to select the most
reliable LABOCA - X-ray counterparts and identify any possible chance
encounters, we use the mid-infrared emission from IRAC and MIPS, whose images
have a much higher source density than those of LABOCA and \chandra.

We first cross-match the positions of the LABOCA sources with the positions of
the 24\,$\mu$m {\it Spitzer} MIPS sources. We choose this band as it is
expected that there will be a correlation between the sub-mm emission and that
at mid-IR wavelengths  (Pope et al. 2006). We use the likelihood ratio (LR) method
(Sutherland \& Saunders 1992) to select the most probable counterpart, choosing the
optimum likelihood ratio threshold that maximizes the sum of reliability and
recovery fraction (see Luo et al. 2010, Rovilos et al. 2010). Starting with an initial
search radius of 8\arcsec and with $LR>0.4$ we find MIPS counterparts for 87\%
of the LABOCA sources which fall in the area mapped by FIDEL, with a mean
reliability of 70\%. Among the LABOCA - MIPS associations are 18/20 LABOCA
sources with an X-ray counterpart. We then check the positions of the X-ray
sources on the FIDEL map to look for any X-ray sources which are associated
with a 24\,$\mu$m source which is not the most probable counterpart to the
LABOCA source; we find one such case, which we remove from our sample. We
repeat the above procedure using the IRAC catalogue and map (from the SIMPLE
survey) to investigate the two LABOCA - X-ray sources with no MIPS association. We
find that the IRAC source associated with the X-rays is not the most probable
counterpart to the LACOCA source in both cases, and remove them from our sample.
We also check the positions of the radio (VLA - 20cm ) sources of
 Kellerman et al. (2008) at a flux limit of 43 $\rm mJy$ ($5\sigma$). 
  We find radio counterparts for 5/18 LABOCA - X-ray
associations (40, 57, 67, 108, 114); their radio positions agree with the FIDEL
positions within 1.4\arcsec.

\begin{figure*}
 \begin{center}
\includegraphics[width=12.0cm]{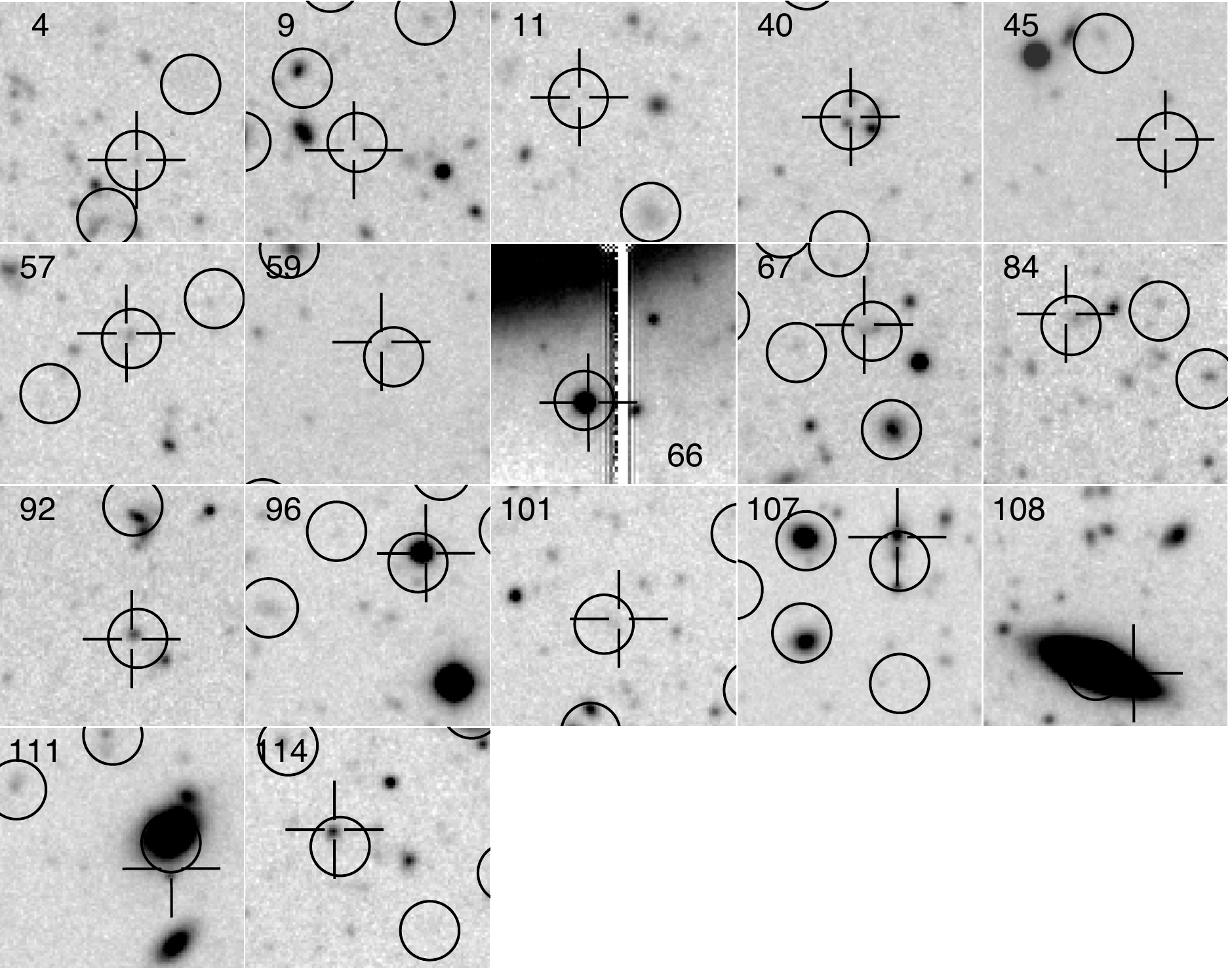} 
\caption{The MUSYC R-band images. The circles (3 arcsec radius) show the MIPS positions while the crosses 
 show the X-ray positions. The thumbnails are centered on the LABOCA coordinates while their size is 25 arcsec.}
  \label{cutouts}
  \end{center}
\end{figure*}

The final step is to check the optical images. In Fig.\,\ref{cutouts} we
plot the MUSYC images of the 17 LABOCA - X-ray candidates. With circles
(3\,arcsec radius) we mark the MIPS positions, while the X-ray sources are
marked with crosses, prefering the CDFS 2\,Ms positions to the eCDFS,
in cases when a source is detected in both surveys. The images are centred in
the LABOCA positions with a 25\,arcsec size. The number on the top-left
of each image is the LABOCA identification. We can clearly see that there are
two cases where the X-ray source is more likely to be associated with a
different optical source than the one producing the bulk of the 24\,$\mu$m flux: these 
are W-107 and W-111. In the case of W-107, the 24\,$\mu$m flux is coming from three
discrete optical sources and the X-ray source is associated with the one
which lies farther away from the LABOCA position. Therefore it is highly unlikely that the
X-rays and the sub-mm emissions are physically connected. In the case of W-111
the X-ray source is different than the 24\,$\rm \mu$m source and again the X-ray
and the sub-mm emissions are not associated. We remove both sources from our
sample.

We note that the X-ray counterpart  of one source (W-96) 
is flagged as a star in Taylor et al. (2009). We
check its optical morphology using {\it HST} imaging from the GEMS survey
(Rix et al. 2004) and we confirm that it has an unresolved morphology with a PSF
of  $\sim 0.08 \arcsec$  (diffraction limit in the $z_{850}$ band). 
 The  X-ray to optical flux ratio is low ($\log[f_{0.5-2\,{\rm keV}}/f_R]=-1.8$) confirming 
  the stellar classification (see e.g. Rovilos et al. 2009).  
On these grounds we remove this source from the
final sample. This of course does not suggest that this is a  star having sub-mm emission,
but simply that the X-rays are related to a stellar source, while there is no
association of the sub-mm source with an X-ray extragalactic source.

Our final sample consists of 14 sources. Ten of them are in the region observed
by the 2\,Ms CDFS and four are in the shallower eCDFS region. The mean reliability
of the sub-mm - MIPS associations for these 14 sources is 87.5\%, i.e. 
 we expect 1.7 spurious sources in our sample.

\section{Photometry and redshifts}

For the photometry in the optical bands we use the MYSYC catalogues of
Taylor et al. (2009) and  Gawiser et al. (2006).  Taylor et al. (2009) provide
photometry of $\sim 17000$ sources of the eCDFS detected in the (CTIO-4\,m -
ISPI) $K$-band in the $U-B-V-R-I-z'-J-H-K$ bands, and Gawiser et al. (2006) use
the same optical data to detect sources in the combined $B-V-R$ band and provide
photometry in the $U-B-V-R-I-z'$ bands.

The {\it Spitzer} photometry comes from the SIMPLE and FIDEL surveys. In the public
SIMPLE catalogue the aperture fluxes are given, and we use the 2\arcsec\ 
apertures and standard aperture corrections derived from the IRAC
cookbook\footnote{http://ssc.spitzer.caltech.edu/irac/iracinstrumenthandbook/}
to calculate the total fluxes. In the FIDEL case we use the corrected isophotal
fluxes derived from the {\sl SEXTRACTOR} source detection code (Bertin \& Arnouts 1996).
We identify as a source an
accumulation of 6 pixels with flux higher than 1.5 times the local rms.
 We calibrate the derived fluxes using the total fluxes of
common sources between the FIDEL and MIPS-GOODS surveys. We also visually
inspect the apertures derived from {\sl SEXTRACTOR} to check for source blending
problems, and also to check if they include all the flux of the source, since the
MIPS PSF has prominent side-lobe features. We find only one such case (W-108), 
 where the aperture used does not encompass all the flux and we consider
the measured flux as a lower limit.

Three sources in our sample (W-40, W-45, and W-67) fall in the area of GOODS which
has a deeper coverage with IRAC and MIPS, as well as the {\it HST}. For these sources
we use the photometry of Grazian et al. (2006) which includes optical and
IRAC bands. These three, along with sources 9 and 11 fall in the MIPS-GOODS area with
deeper coverage than FIDEL, and therefore we use the GOODS $24\,\mu{\rm m}$ photometry.

As a final step, we inspect the images of the 14 LABOCA - X-ray sources in all
the bands used and compare them with with those with the smallest PSF; the
optical GEMS images and (where available) the GOODS-{\it HST} images. We do that in
order to detect whether the flux measured in the bands with the larger PSFs are
blended with nearby sources. We find that sources W-59 and W-67 are in fact double
sources blended in the IRAC and MIPS images because of limited spatial
resolution, whereas for sources 9, 40, and 84 there are nearby sources which
affect the MIPS flux. Source 67 is in the GOODS area and has optical and IRAC
photometry from the GOODS-MUSYC survey Grazian et al. (2006), which uses PSF
matching to the {\it HST} images to measure the fluxes in the optical and infrared
bands, so we consider the problem to be confined only in the MIPS photometry
(taken from GOODS).

 Four sources (W-57, W-67, W-108,
and W-114) have spectroscopic redshifts assigned to their optical, infrared, or
radio counterparts:
Szokoly et al. (2004), Kriek et al. (2006), Norris et al. (2006), Chapin et al. (2010) respectively.
The photometric  redshifts in the area of the CDFS come from
Luo et al. (2010), where photometric redshifts are calculated for X-ray
sources using up to 42 UV, optical, and infrared bands. The photometric redshifts 
 in the area of the eCDFS come from Taylor et al. (2009). Finally,  
the photometric redshift of source  W-4 in the eCDFS is calculated using the  EAZY (Brammer et al. 2008) code
 using the $B$, $V$, $R$, $I$, and $z'$ bands. This source does not have a photometric redshift in 
  the K-selected sample of Taylor et al. (2009) 
  as it has not been detected in this band.   
In Table\,\ref{photometry} we present the X-ray, optical, and infrared fluxes as well as the available redshifts 
of the sources in our sample.

\begin{table*}
\centering
\caption{Photometry}
\label{photometry}
\begin{tabular}{ccccccccccc}
\hline\hline
ID &  $\alpha$ & $\delta$   & $f_{0.5-10\,keV}$ & $R$ & $f_{3.6\,\mu{\rm m}}$ & $f_{24\,\mu{\rm m}}$ & $f_{850\,\mu{\rm m}}$ & $LR$ & z & Ref. \\
  (1)    &    (2)      & (2)  &   (3)  &   (4)     &  (5)    & (6)   &  (7) & (8) & (9)  & (10)  \\
\hline
 W-4 & 52.8996 & -27.9121 &  1.85 & 24.94 &   $7.00\pm 0.10$   &    $139\pm 5$   & $11.0\pm 1.2$ &  5.00  &  $2.58\pm 0.9$  &  1 \\
 W-9 & 53.0475 & -27.8704 &  7.14 & 26.02 &   $9.83\pm 0.05$   &    $122\pm 5$ $^b$ &  $9.4\pm 1.2$ &  6.92 & $3.99\pm0.08$  & 2 \\
W-11 & 53.0577 & -27.9334 &  1.33 & 25.61 &   $6.83\pm 0.07$   &    $113\pm 6$   &  $9.2\pm 1.2$ &  5.06  & $6.07\pm 1.1$  & 2 \\
W-40 & 53.1950 & -27.8557 &  0.44 & 24.35 &   $8.51\pm 0.15$   &    $140\pm 4$ $^b$ &  $6.4\pm 1.2$ &  6.08 & $1.90\pm 0.02$  & 2 \\
W-45 & 53.1052 & -27.8752 &  0.06 & 27.41 &   $7.50\pm 0.27$   &    $141\pm 4$   &  $6.3\pm 1.2$ &  1.50 & $2.50\pm 0.2$  & 2 \\
W-57 & 52.9665 & -27.8908 &  1.51 & 24.33 &   $7.60\pm 0.10$   &    $273\pm 6$   &  $6.1\pm 1.3$ & 12.81 & 2.940 & 3 \\
W-59 & 53.2658 & -27.7362 &  1.35 & 25.62 &  $30.28\pm 0.10$ $^a$ &  $453\pm 5$ $^a$ &  $6.0\pm 1.3$ &  7.67  & $1.38\pm 0.02$  & 2 \\
W-66 &   53.3830 & -27.9028  &   36.1    & 20.72  &  $56.34\pm0.13$  &  $679 \pm9.4$   &  $6.1\pm 1.3$  & 5.37      & $0.58\pm 0.03$ & 5 \\
W-67 & 53.1803 & -27.9206 &  0.29 & 24.20 &   $21.7\pm 0.9$    &    $554\pm 4$ $^a$ &  $5.9\pm 1.3$ & 15.77 & 2.122 & 4 \\
W-84 & 52.9773 & -27.8513 &  1.58 & 24.97 &  $12.98\pm 0.09$   &    $231\pm 4$ $^b$ &  $5.5\pm 1.3$ &  4.16 & $3.2 \pm 0.06$ & 2 \\
W-92 & 52.9096 & -27.7277 &  2.98 & 24.49 &  $26.57\pm 0.10$   &     $80\pm 5$   &  $5.2\pm 1.2$ &  2.69 & $2.10\pm 0.04$  & 5 \\
W-101 & 52.9643 & -27.7649 &  1.16 & 23.01 &   $5.33\pm 0.10$   & $34\pm 2.9$   &  $5.1\pm 1.3$ &  0.66 & $2.53\pm 0.12$ & 2 \\
W-108 & 53.3175 & -27.8445 &  1.75 & 17.36 & $516.74\pm 0.10$   & $4040\pm 100$ $^c$ &  $5.0\pm 1.2$ &  7.89 & 0.0875 & 6 \\
W-114 & 52.9628 & -27.7436 &  3.11 & 23.88 &  $51.63\pm 0.10$   &    $660\pm 9$   &  $4.9\pm 1.3$ &  2.85 & 1.605 & 7 \\
\hline\hline
\end{tabular}
\begin{list}{}{}
\item The columns are:   (1) LABOCA source number in Wei\ss et al. (2009); (2) X-ray counterpart Equatorial coordinates; 
(3) Total band X-ray flux in units of ${\rm \times 10^{-15}\,erg\,s^{-1}cm^{-2}}$; (4) R-band magnitude (AB) ; (5) 3.6$\rm \mu m$ flux in units of $\rm \mu Jy$; (6) 24$\rm \mu m$ flux
 in units of $\rm \mu Jy$ ; notes: (a) Double source 
blended in the IRAC and/or MIPS PSFs; (b) nearby sources possibly affecting the measured flux
 (c) Lower limit  (7)  LABOCA 870 $\rm \mu m$ flux in units of $\rm mJy$(8) Likelihood-ratio between LABOCA and MIPS counterpart 
 (9) Redshift: Three and two decimal digits denote a spectroscopic 
  and a photometric redshift respectively (10)  
Redshift reference:  1 EAZY; 2 Luo et al. (2010); 3 Szokoly et al. (2004); 4 Kriek et al. (2006); 5 Taylor et al. (2009); 
 6 Norris et al. (2006); 7 Chapin et al. (2010) 
 \end{list}
\end{table*}

\section{X-ray Properties} 

The X-ray properties can give a first idea on the nature of the X-ray detected SMGs. 
 X-ray luminosities higher than about a few times $10^{42}$\,\lunits are usually 
  attributed to AGN activity.
   This limit has been dictated by the highest X-ray luminosity observed in  local
    star-forming galaxies (e.g. Zezas, Georgantopoulos \& Ward 1998,
   Moran, Lehnert \& Helfand 1999).   However, given that SMGs 
    are probably the most powerful star-forming systems it is 
     not unlikely that their X-ray luminosity may surpass this limit. 
   Star-forming systems present very little, if any obscuration 
    (e.g. Georgakakis et al. 2006, 2007, Tzanavaris et al. 2006, Rovilos et al. 2009 and references therein). 
     Therefore the detection 
     of a substantial obscuring column, through X-ray spectroscopy  
     in a low X-ray luminosity source can further differentiate 
      between a star-forming galaxy and an AGN. 
     
     \subsection{X-ray spectroscopy} 
We use the {\sl SPECEXTRACT} script in the CIAO v4.2 software 
 package to extract the spectra  of the 14 X-ray sources in our sample. 
  The extraction radius varies between 2 and 4 arcsec with increasing off-axis angle. 
   At low off-axis angles ($<$4 arcmin)  this encircles 90\% of 
 the light at an energy of 1.5\,keV.  
 The same script extracts response and auxiliary files. 
  The addition of the spectral, response and  auxiliary files has been 
   performed with the FTOOL  tasks {\sl MATHPHA}, {\sl ADDRMF} and {\sl ADDARF} respectively.  
    We use the C-statistic technique (Cash 1979) specifically developed
to extract spectral information from data with low signal-to-noise ratio. 
 We use the XSPEC v12.5 software
package for the spectral fits (Arnaud 1996).  

We fit the data using a power-law absorbed by a cold absorber.  
 We treat both the intrinsic column density 
and the photon index ($\Gamma$)  as free parameters.
 However, in five cases the photon statistics are  low, not allowing us 
  to derive meaningful constraints for both $N_{\rm H}$ and $\Gamma$.   
       In these cases  we can only fix 
        the power-law photon index to $\Gamma=1.8$
 (Nandra \& Pounds 1994, Tozzi et al. 2006), 
         leaving only the column density $N_{\rm H}^{\rm eff}$ as the free parameter
          and vice-versa i.e. setting $N_{\rm H}=0$ and leaving 
           the photon index $\Gamma^{\rm eff}$ as a free parameter. 
         The spectral-fit results are presented  in Table\,\ref{xtable}.
            As it is customary in X-ray Astronomy, errors
 correspond to the 90\% confidence level. 
  
 \subsection{AGN vs. star-forming galaxies} 
\begin{figure}
 \begin{center}
\includegraphics[width=8.0cm]{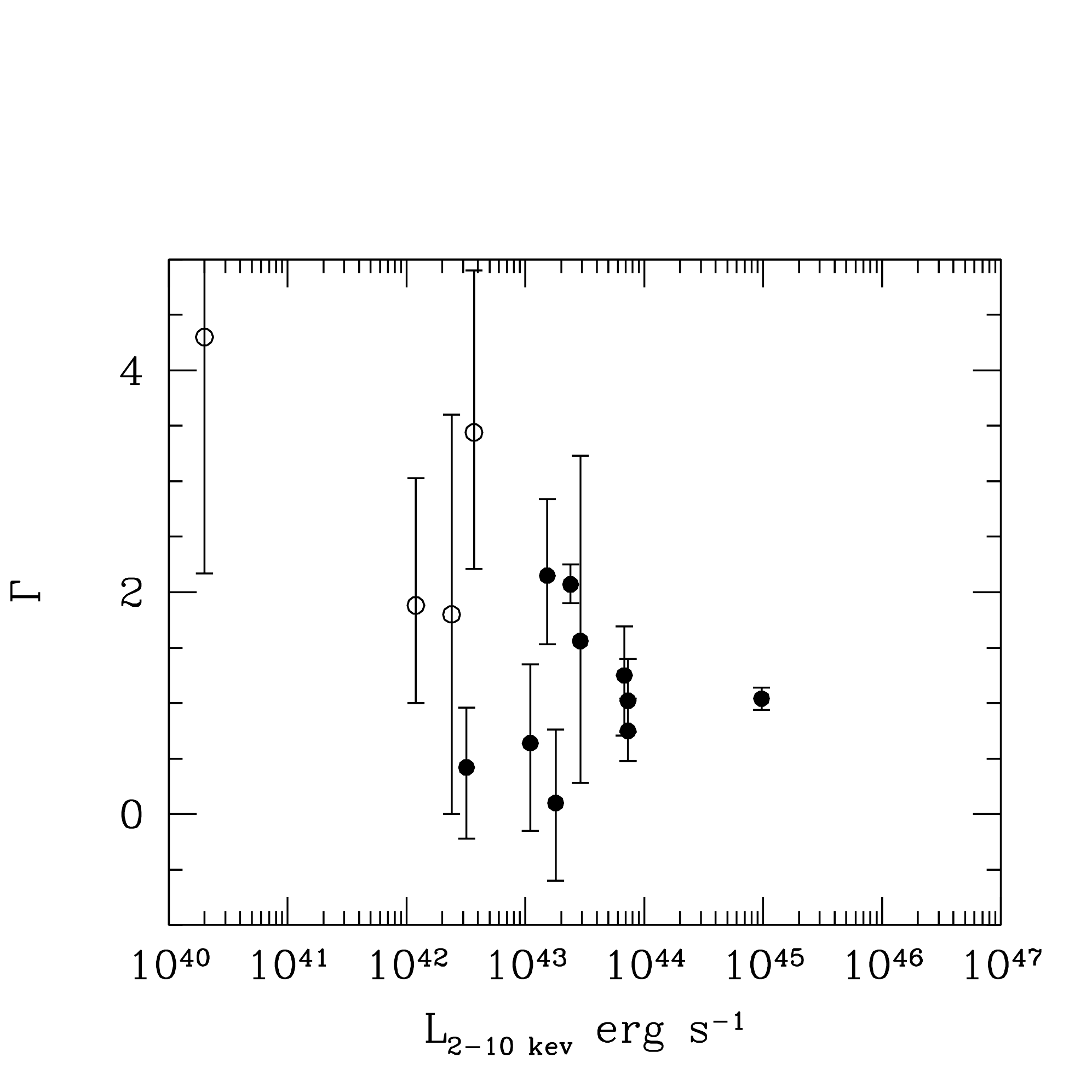} 
\caption{The effective photon index as a function of X-ray luminosity. Filled and 
 open circles correspond to AGN and galaxies respectively (see text).}
  \label{gamma}
  \end{center}
\end{figure}

  In Fig.\,\ref{gamma}  we plot the effective photon index 
         as a function of the  obscured X-ray luminosity, 
         i.e. column (7) vs. column (8) in Table\,\ref{xtable}. 
    We see that nine  sources 
    have high X-ray luminosities ($>10^{43}$ \lunits) 
   and thus can be considered as bona-fide AGN. 
    Four out of the remaining low luminosity sources 
     present soft spectra and thus are possibly  
      associated with star-forming galaxies. 
       The remaining source (W-114) has a hardish spectrum and thus 
        could be a highly obscured AGN: its spectrum 
         is marginally intrinsically  flat ($\Gamma <1.32$ at the 90\% 
        confidence level)  and thus  possibly 
        associated with a Compton-thick AGN with a 
         reflection dominated spectrum.
          
    \subsection{X-ray obscuration}
\begin{figure}
 \begin{center}
\includegraphics[width=8.0cm]{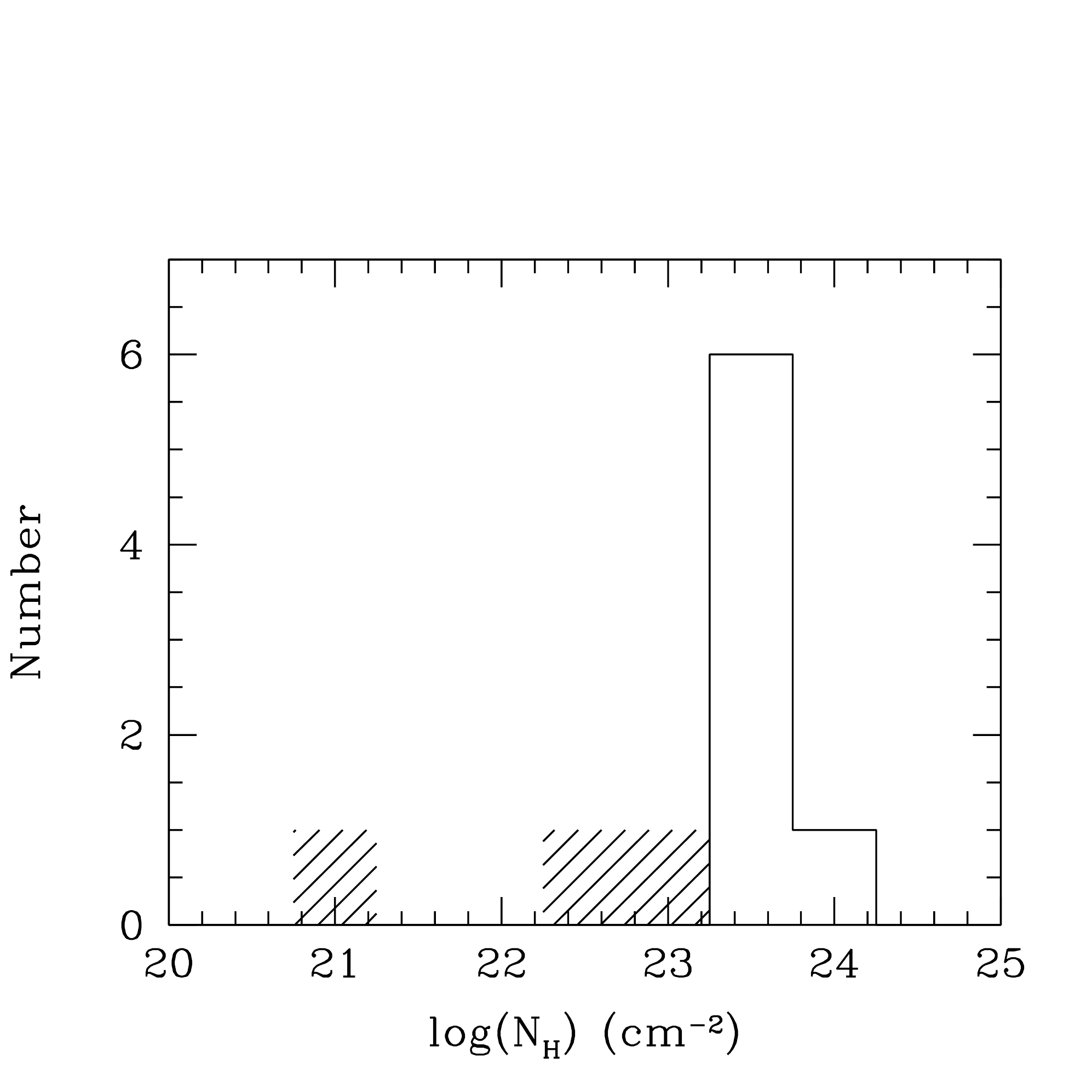} 
\caption{The rest-frame column density distribution for  the ten AGN (see text for details).  The open and hatched histograms correspond to 
 the detections and upper limits respectively.}
  \label{nh}
  \end{center}
\end{figure}

 In Fig.\,\ref{nh} we plot the distribution of the rest-frame column densities
  (column 6 in table 2). 
        It is evident that most AGN (7/10) present column densities exceeding 
       $\rm 10^{23}\,cm^{-2}$, i.e. they are characterised as heavily obscured AGN.         
 One source could be a transmission dominated Compton-thick AGN (W-11).  
    However, the uncertainty in the photometric redshift certainly makes this classification dubious.  
    
\begin{table*}
\begin{center}
\caption{X-ray spectral fits}
\label{xtable}
\begin{tabular}{lccccccc}
\hline \hline
ID & XID & z &  $N_H$      & $\Gamma$       & $N_H$  ($\Gamma=1.8$)   &    $\Gamma$ ($N_H=0$) & $ L_X$  \\ 
(1)   & (2)&    (3)        &  (4)          &  (5)         &   (6)     & (7)  & (8) \\
\hline
  W-4    & LE-97                 &  2.58       & $<15.3$           &  $1.55^{+2.14}_{-1.24}$ & $<15.0$  &  $1.56^{+1.67}_{-1.28}$      & 2.9 \\
  W--9    & LU-114  LE-319    &  3.99   & $28.3^{+8.5}_{-7.8}$ & $1.72^{+0.16}_{-0.19}$  & $31.1^{+4.2}_{-4.2}$ & $1.04^{+0.10}_{-0.10}$ & 97.\\ 
  W-11   & LU-131 , LE-332    &  6.07 & $196^{+139}_{-93}$  &   $2.32^{+0.47}_{-0.67}$        &   $171^{+115}_{-74}$ &  $1.02^{+0.38}_{-0.28}$ &   7.3 \\
  W-40  &  LU-385                 &    1.90    &  -                                   &   -      &   $21^{+14.8}_{-13.6}$  & $0.64^{+0.71}_{-0.79}$  & 1.1  \\
  W-45$\dagger$  &  LU-230                 &    2.50 &  -                                      &  -       &  $<5.53$                              &  $3.44^{+1.46}_{-1.23} $ &  0.37 \\
  W-57  & LU-26,  LE-203    &2.940 & $39.4^{+32.9}_{-26.5}$    &  $2.60^{+1.18}_{-1.01}$       &  $20.1^{+15.9}_{-12.9}$    &  $1.25^{+0.44}_{-0.54}$ & 6.8   \\
  W-59$\dagger$  &  LU-441                 &  1.38  &  -                                     &   -       &   $<6.1$    & $1.88^{+1.15}_{-0.88}$ &  0.12 \\
  W-66    &  LE-725 &          0.58          &  $<0.17$   & $2.04^{+0.25}_{-0.17}$ & $<0.06$  & $2.07^{+0.18}_{-0.17}$  & 2.4 \\
  W-67$\dagger$  &  LU-362                 &   2.122 &  -                                   &  -       &       -           &     1.8                                 &  0.24     \\
  W-84  &  LU-33                   &    3.20   & $<30.5$                       &  $1.03^{+0.68}_{-0.49}$ &   $26.9^{+16.1}_{-7.6}$   &  $0.75^{+0.29}_{-0.27}$  & 7.3 \\
  W-92  &  LE-112                 &    2.10     &  $3.6^{+21.0}_{-13.4} $ &    $0.33^{+1.53}_{-0.83}$  &   $34.7^{+25.1}_{-18.0}$          &    $0.10^{+0.66}_{-0.70}$    &   1.80  \\
  W-101 &  LU-25                   &   2.53     &   $<7.9$             &      $2.11^{+1.4}_{-0.6}$         &   $<2.8$                                &   $2.15^{+0.69}_{-0.62} $     & 1.53  \\
  W-108$\dagger$  &  LE-634                &    0.0875  &   -                               &    -                  &    $<0.62$   &        $4.3^{+2.63}_{-2.13}$    &    $2\times10^{-3}$  \\
  W-114 &  LU-23                 &   1.605 &  $<45.0$                         & $-0.15^{+1.47}_{-1.02}$ &  $31.9^{+34.2}_{-12.6}$   & $-0.42^{+0.54}_{-0.64}$  &  0.43 \\
\hline \hline 
\end{tabular}
\begin{list}{}{}
\item The columns  are: (1) Laboca ID in Wei\ss  et al. (2009) (2) CDFS (LU) or eCDFS(LE) ID in Luo et al. (2008)   
 or Lehmer et al. (2005) respectively   (3) redshift; (4),(5) rest-frame column density and photon index  in the case of sources with 
   good photon statistics (see text) (6) rest-frame column density in units of $10^{22}$ \cunits for fixed $\Gamma=1.8$ (7) photon-index $\Gamma$ 
    for column density $\rm N_H=0$ (8) {\bf Intrinsic} X-ray Luminosity in the 2-10 keV band in units of $10^{43}$ \lunits, obtained by setting 
     $\rm N_H=0$ in column (6); 
    $\dagger$ denotes a possible star-forming galaxies on the basis of the X-ray properties i.e. low X-ray luminosity and soft spectrum  (see text). 
\end{list}
\end{center}
\end{table*}

\section{Mid-IR properties}

\subsection{Mid-IR Spectral Energy Distributions}
\begin{figure}
   \begin{center}
\includegraphics[width=8.0cm]{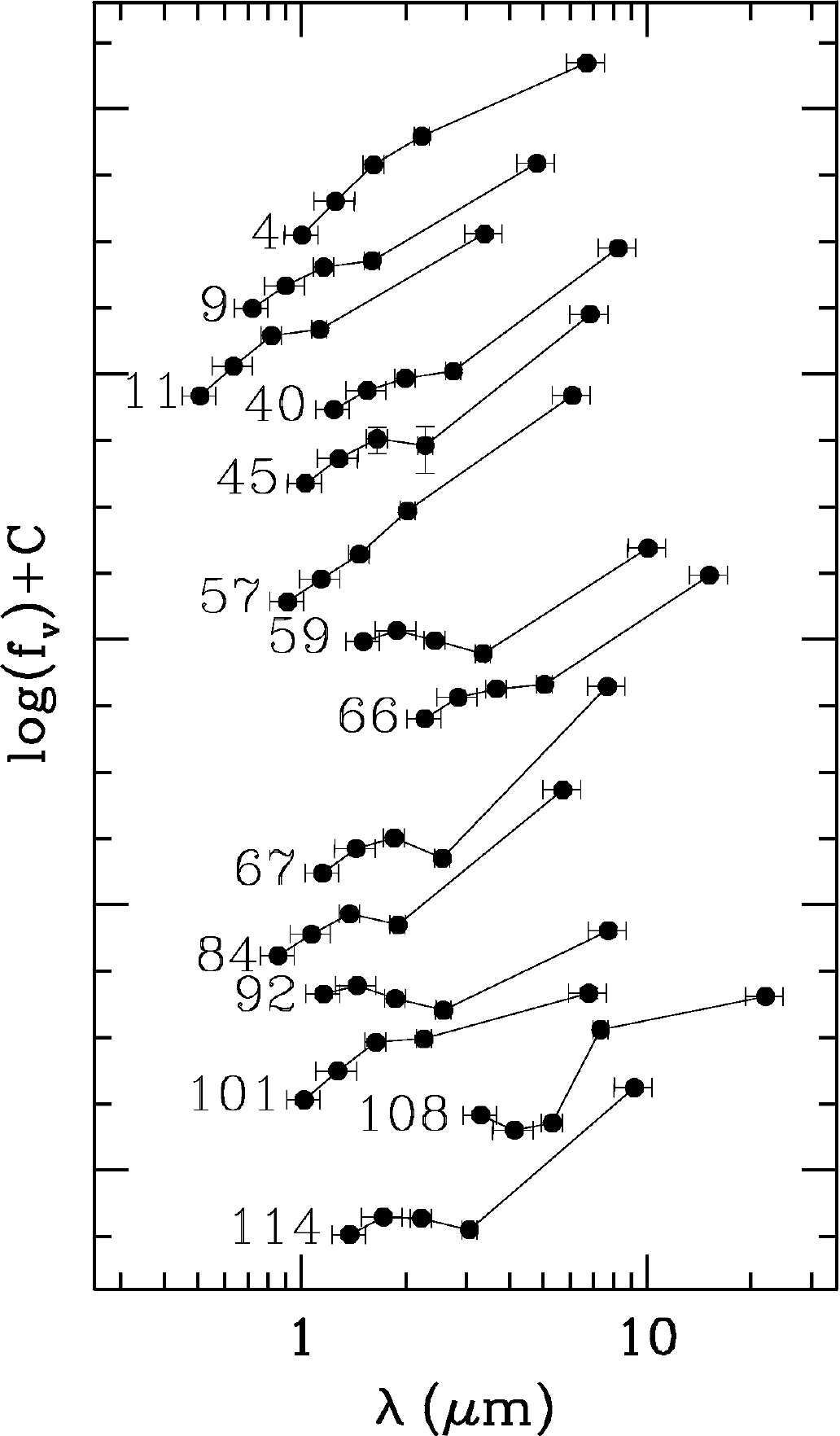} 
\caption{The IRAC, MIPS (24$\rm \mu m$) Spectral Energy Distributions}
  \label{SED}
  \end{center}
\end{figure}

 We construct the spectral energy distribution (SED) 
 in order to obtain an idea on the dominant 
 powering mechanism (AGN or star-formation) in the 
 mid-IR part of the spectrum. The SEDs are also used 
  in order to obtain an accurate estimate of the infrared luminosities.     
  Power-law SEDs in the mid-IR are characteristic of AGN 
   emission (e.g. Alonso-Herrero et al. 2006, Polletta et al.  2007). 
   A prototype power-law mid-IR SED in the local Universe is 
    that of Mkn231 (Armus et al. 2007). This is a broad-absorption-line QSO (Braito et al. 2004) 
     and a  Ultraluminous Infrared IRAS galaxy. 
    On the other hand star-forming sources have a distinct dip in their mid-IR spectra at a 
   rest-frame wavelength below $\rm 10\,\mu m$.  Such a source is Arp\,220 (e.g. Iwasawa et al. 2005), 
    a ULIRG whose far-IR SED is dominated by a very strong star-forming component (Armus et al. 2007).
 We present the IRAC/MIPS SEDs 
 in Fig.\,\ref{SED}.  There are only two clearcut power-law SEDs suggestive of AGN: W-4 and W-57.  

  Apart from this first order approach, we fit a star-forming 
   in combination with a torus template to the SED from $\rm 3.6\,\mu m$ up to $\rm 870\,\mu m $.
   We use the SED templates of Polletta et al. (2007)  in order to produce templates
combining star formation and AGN activity. As star forming templates we use
those of M\,82 and Arp\,220. These are two actively star forming galaxies
differing in the star formation rate and the amount of dust. Arp\,220 is
producing stars with a rate of $270\,{\rm M_{\odot}\,yr^{-1}}$
 (Shioya et al. 2001) and its interstellar dust gives it an infrared luminosity
$>10^{12}{\rm L}_\odot$, making it the prototypical ULIRG. M\,82 on the other
hand has a more modest star formation rate of $10\,{\rm M_{\odot}\,yr^{-1}}$ and
lower mid-to-far infrared luminosity. The two SEDs are almost identical in the
near-to-mid infrared and differ significantly in their mid infrared PAH and
silicate features and the mid-to-far infrared dust emission. As an AGN-only SED
we use the Torus SED of  Poletta et al. (2007), which is a fit to the SED of a
heavily obscured type-2 QSO, SWIRE\,J104409.95+585224.8  (Polletta et al. 2006).
We start the SED fitting by choosing the best linear combinations of
Arp\,220+Torus and M\,82+Torus templates using the four IRAC and the 24\,$\mu$m
datapoints and minimizing the $\chi^2$ value of the fit. Then we calculate the
$\chi^2$ values of the optimum combinations using the previous datapoints plus
the sub-mm flux and choose the combination (Arp\,220+Torus or M\,82+Torus)
which minimizes this value. In other words we use the mid-infrared to determine
the relative AGN and starburst contribution (where the two starburst templates
are almost identical) and define the optimum starburst template by its
prediction of the far-infrared flux. Then, we use the F-test to determine
whether the AGN contribution is needed to fit the mid-infrared flux; we
calculate:
\[F=\frac{\frac{\chi^2_1-\chi^2_2}{p_2-p_1}}{\frac{\chi^2_2}{n-p_2}}\]
where $\chi^2_1$ and $\chi^2_2$ are the best-fit $\chi^2$ values of the
starburst-only and combination templates respectively, $p_1$ and $p_2$ are the
degrees of freedom of the two cases (here $p_1=1$ and $p_2=2$) and $n$ is the
points used to make the fit (here $n=5$). We assume that the AGN contribution is
needed if the false-rejection probability is lower than 10\% ($F>8.5$).

\begin{figure*}
   \begin{center}
\includegraphics[width=12.0cm]{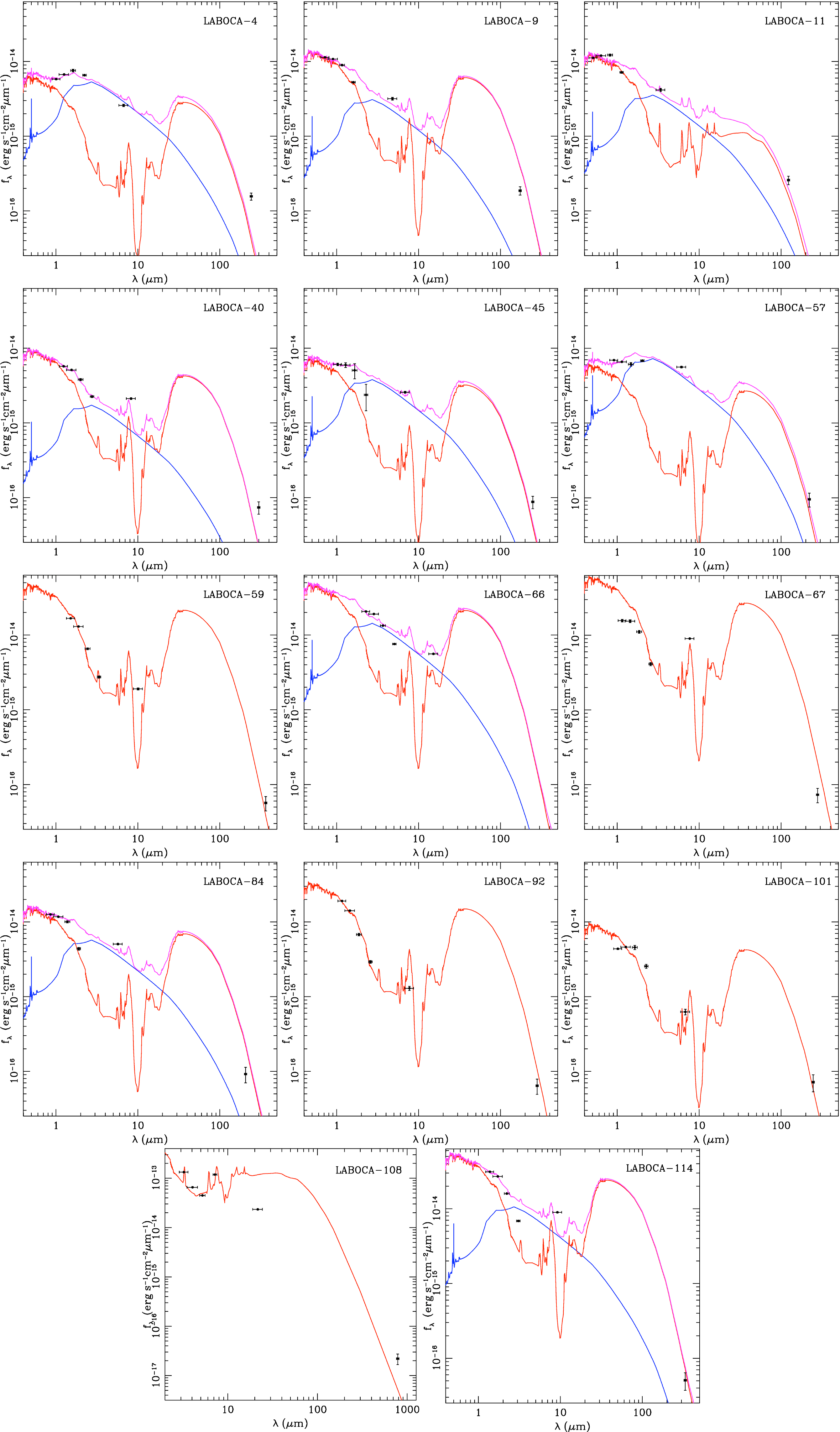} 
\caption{IR through sub-mm Spectral Energy Distributions. The black points 
denote the observations while the solid lines the fitted templates:
 the red and blue lines represent the star-forming and torus templates of 
  Polletta et al. (2007) while the purple curve denotes the total model}
  \label{SED_full}
  \end{center}
\end{figure*}

The template fits are presented in Fig.\,\ref{SED_full}, while 
  the derived IR luminosities are given in Table\,\ref{lir}. 
  We note that the IR-luminosity is estimated from the
mid-infrared SED and  the $\rm 870\,\mu m$ flux. This does not
probe the peak of the infrared SED (around rest-frame 70$\rm \mu m$ in $\nu F_\nu$) and thus introduces an
appreciable uncertainty in the determination of the far-IR luminosity. 
Measurements of the infrared SED at its peak (for example with {\it Herschel}) 
are needed in order to provide accurate bolometric luminosities. 
However, we note that the $\rm 6\,\mu m$  luminosities are hardly affected
 by the lack of far-IR measurements.
   We see that a torus component is required in most cases (eight out of ten  AGN). 
    However, the contribution of the torus to the total IR luminosity is small (see Table\,\ref{lir}).  
   The contribution of the torus is appreciable only in the case of the two power-law IRAC AGN (W-4 and W-57)
    and also in the case of the source W-11  (reaching up to $\sim 20\%$). 
    For most candidate galaxies the bulk of the $\rm 6\mu m$ emission comes from star-forming processes.  
     Only the candidate galaxy W-45 has a non-negligible torus component ($\sim 10\%$). 
         
\begin{table*}
 \begin{center}
 \caption{Infrared Luminosities}
 \label{lir}
 \begin{tabular}{lllllllll}
 \hline \hline
  ID      &   $\rm \nu L_\nu(6\mu m)$  &  $L_{IR}$ & $L_{torus}/L_{IR}$ & $\rm \log( L_X/L_{IR})$ & IRAC & Full-SED & Prob. \\
  (1)     &   (2)                  &       (3)             &       (4)         &        (5)      &       (6)     &  (7)   & (8) \\
 \hline
 W-4  &  1.0       &  1.4 & 0.14 & -2.7 & AGN  & Arp220/Torus & 0.5 \\
 W-9   & 1.7 & 7.8 & 0.04  & -1.9 & SF & Arp220/Torus & 3.9\\           
W-11  & 5.4  &  4.7 & 0.21 & -2.8 & SF & M82/Torus & 2.1 \\     
W-40  &  0.16  &  0.88 & 0.03&  -2.9 & SF & Arp220/Torus & 0.4\\        
W-45$^\dagger$  & 0.69  &   1.4 & 0.09& -3.6 &  SF & Arp220/Torus & 0.9 \\     
W-57  & 2.0  &  1.9 & 0.18 & -2.4 & AGN & Arp220/Torus & 0.6 \\       
W-59$^\dagger$  & 0.06$^\star$  &  1.7 & 0.006 & -4.15 & SF&  Arp220  & 42. \\
W-66 &   0.07   &    0.23 &  0.05 &  -1.98 & SF & Arp220/Torus & 3.9 \\
W-67$^\dagger$  & 0.76$^\star$  &  4.2 & 0.03 & -4.24 & SF & Arp220 & 47.  \\     
W-84  & 1.9  & 5.3 & 0.07 &  -2.86 & SF &  Arp220/Torus & 2.7 \\      
W-92  & 0.0$^\star$  & 3.8  & 0.00 & -3.30& SF &  Arp220 & 99. \\      
W-101 & 0.22$^\star$  & 1.30 &  0.03 &  -2.92 & SF &  Arp220 & 11.1  \\     
W-108$^\dagger$ & 0.002$^\star$ & 0.02 & 0.03 & -3.84 & SF &  M82 & 17.1 \\         
W-114  & 0.65$^\star$ &  3.3 & 0.03 &  -4.01 & SF & Arp220/Torus  & 5.2 \\ 
   \hline \hline
 \end{tabular}
\begin{list}{}{}
\item The columns  are: (1) LABOCA ID as in Table\,\ref{photometry}
   (2) Torus 6$\rm \mu m$ $\nu L_\nu$ monochromatic luminosity in units of  $10^{45}$ \lunits.  
   (3) 10-1000 $\rm \mu m$ Infrared Luminosity in units of $10^{46}$ \lunits. 
   (4) Ratio of torus to total 10-1000 $\rm \mu m$ luminosity 
   (5) Logarithm of the Ratio of X-ray luminosity to total IR (10-1000 $\rm \mu m$) luminosity. 
   (6) mid-IR IRAC template: power-law (AGN) vs. curved (star-forming) 
   (7) Best-fit templates from Polletta et al. (2007). (8) Probability ($\times 10^{-2}$) that the torus component 
    is not needed according to the F-test. Notes: ($\dagger$) denotes a possible 
    star-forming galaxy on the basis of the X-ray properties; ($\star$) means that the best-fit 
     value of the torus luminosities are given although such component is not formally required by F-test.
\end{list}
 \end{center}
 \end{table*}

 \subsubsection{X-ray to $\rm 6\,\mu m$ luminosity} 
  The derivation of the 
   $L_{\rm x}$ to $L_{\rm 6\,\mu m}$ ratio has as a goal to
    identify any highly obscured, Compton-thick AGN.  
     These are difficult to identify with X-ray spectroscopy 
      especially in the case of limited photon statistics.   
    One of the  most reliable proxies of the intrinsic power of an 
 AGN is the mid-IR monochromatic $\rm 6\,\mu m$ luminosity (e.g. Lutz et al. 2004). 
  This wavelength region is more representative 
 of the hot dust heated by the AGN and thus provides a 
  reliable diagnostic of the AGN power (e.g. Alexander et al. 2008).  
       
\begin{figure*}
   \begin{center}
\includegraphics[width=12.cm]{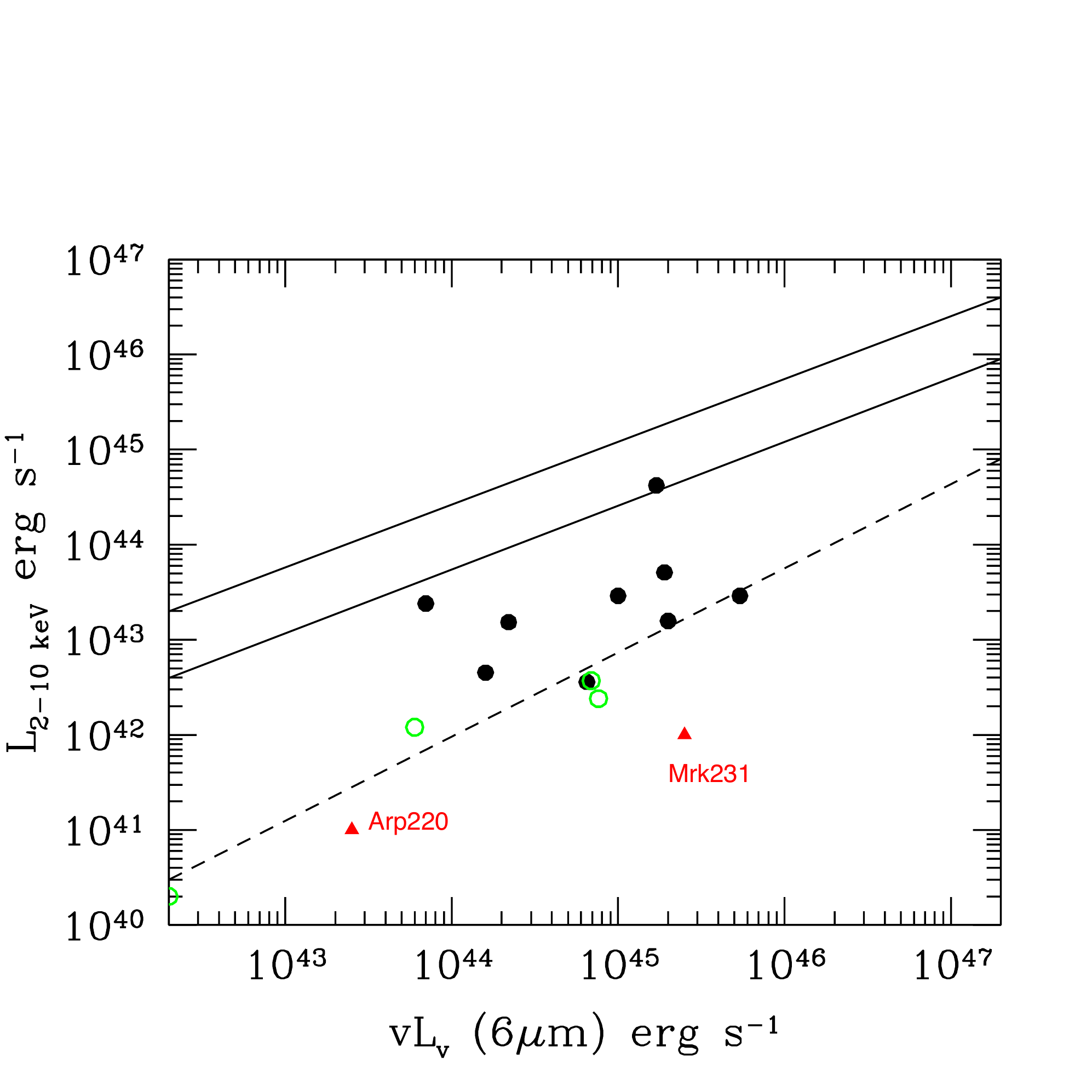} 
\caption{The  X-ray (2-10 keV)  luminosity (uncorrected for absorption) as a function of the torus 6$\rm \mu m$ 
 luminosity. Filled (black) circles represent the AGN while open (green) circles denote 
  the candidate galaxies. Note that only nine AGN are plotted as source W-92
   shows no torus contribution.  
  The solid lines denote the region occupied by unabsorbed 
 AGN while the region beyond the short dash line is populated by Compton-thick 
 AGN in the local Universe (adapted from Alexander et al. 2008).
 For comparison we plot the positions of Arp220 and the Compton-thick AGN Mrk231 
 on the diagram.}
  \label{lxl6}
  \end{center}
\end{figure*}

 We present the 2-10\,keV absorbed luminosity  against the 
 monochromatic $\rm 6\,\mu m$ IR  luminosity coming from the torus in Fig.\,\ref{lxl6}.
 The area between the solid lines denotes the region  
  of the X-ray to $\rm 6\,\mu m$ luminosity plane where local AGN reside (Lutz et al. 2004, 
   Alexander et al. 2008, Bauer et al. 2010).
  The area below the dashed line corresponds to low X-ray 
 luminosity sources, i.e. Compton-thick sources (or alternatively normal galaxies).      
   We see that some sources fall in the low part of the $L_{\rm x}/L_{\rm 6\,\mu m}$ diagram (e.g. W-11, W-57, W-114). 
   All four candidate galaxies, i.e. the sources with low X-ray luminosities and soft X-ray spectra
    fall as well in the low part of this diagram.

\begin{table*}
 \begin{center}
 \caption{X-ray Stacking Results}
\label{stacking} 
\begin{tabular}{lllllll}
 \hline \hline
  Sample           &No        &  Soft Counts     & Soft Flux & Hard Counts & Hard Flux & HR  \\
  (1)                   & (2)       &   (3)                  & (4)           &   (5)             &  (6)            &  (7) \\
  \hline
LABOCA eCDFS             & 100       &  61 (5.6$\sigma$) &  $2.3\times10^{-17}$  &  26 (2.2$\sigma$) & $2.2\times 10^{-17}$ & $-0.40\pm0.10$ \\
LABOCA CDFS              & 21         &   28 (2.5$\sigma$)  &   $5\times 10^{-18}$   &  16 (1.3$\sigma$) & $<1.6\times 10^{-17}$  & $>-0.06$$^\dagger$   \\
  \hline \hline
 \end{tabular}
 \begin{list}{}{}
\item (1), (2) sample where stacking technique has been applied and number of sources. (3) Net soft counts and significance of signal. 
 (4) Soft flux (0.5-2 keV) in units of \funits. (5) Hard net counts and significance. (6) Hard flux (2-5 keV)  in units of \funits. (7) Hardness ratio. 
 ($\dagger$):  the upper limit on flux and the corresponding lower limit on the hardness ratio correspond to $2\sigma$.  
\end{list}
 \end{center}
 \end{table*}

\section{X-ray Stacking Analysis}
Having dealt with the properties of the X-ray detected SMGs, we 
 attempt to gain an insight on the  average X-ray properties of the 
  individually undetected LABOCA sources. This can be done in a 
   statistical manner simply by co-adding (stacking) the X-ray flux around each SMG. 
Stacking techniques have been widely used in X-ray Astronomy to study the mean properties of source populations selected to have certain 
well defined properties and which are too faint in X-rays to be individually detected (e.g. Nandra et al. 2002; Georgantopoulos et al. 2008).
A fixed radius aperture is used to extract and to sum the X-ray photons at the positions of the LABOCA sources. Sources that lie close
 to or are associated with an X-ray detection are excluded from the analysis to avoid contamination of the stacked signal from the X-ray photons of detected sources. 
 In order to maximise the signal we adopt an extraction radius of 3 arcsec around the MIPS $\rm 24\,\mu m$ counterparts of the LABOCA sources.  
  A 3\,arcsec aperture encloses more than about 90\% of the photons in the
 soft 0.5-2\,keV spectral band at an off-axis angle of 5\,arcmin.
  We perform the stacking analysis in two energy bands: the soft (0.5-2\,keV) and the hard  (2-5\,keV).
   The choice of the latter band minimises the instrumental background contamination thus maximising the sensitivity of the stacking analysis.
   We perform the stacking analysis for the LABOCA sources lying in both the eCDFS and CDFS regions.
    
  The significance of the stacked signal depends on the value 
of the background. This is estimated by using the background maps produced by the WAVDETECT task of CIAO.  We 
sum the X-ray photons in regions around each source used
in the stacking.  The significance of the stacked signal in background 
   standard deviations is estimated by (T - B) / T, where T and B are the total (source + background) and background counts respectively.
Fluxes are determined by multiplying the stacked count rate by the appropriate energy conversion factor, 
 which is estimated separately for each class of sources, based on the spectral shape of the stacked signal. 
  A rough estimate of the mean spectral index is obtained by deriving the hardness ratio  (HR=(h-s)/(h+s)) 
   between the s=0.5-2\,keV and h=2-5\,keV bands.  
 
  The stacking results are summarised in Table\,\ref{stacking}. 
 We see that the stacking of the LABOCA sources in the eCDFS  reveals a strong signal in the soft band while the signal is marginal 
  in the hard band. We have checked the significance of the soft signal 
   by performing 100 simulations around random positions on the image. 
    There is no simulation where we obtain a signal  higher than the actually detected signal (61 counts), 
     implying that the confidence limit for our detection is higher than 99\%. 
    The implied flux in the soft band is $\rm 2.3\times10^{-17}\,erg\,cm^{-2}\,s^{-1}$, i.e. about four times 
   fainter than the flux limit in the eCDFS (see Virani et al. 2008, Lehmer et al. 2005). 
    We find a hardness ratio of $-0.40\pm 0.10$  corresponding to a photon index of $\Gamma\approx1.6$. 
     We note that in the CDFS we barely detect a signal even in the soft band. This is most probably because 
       of the small number of SMGs employed in the stacking analysis.

 \begin{figure*}
   \begin{center}
\includegraphics[width=9.cm]{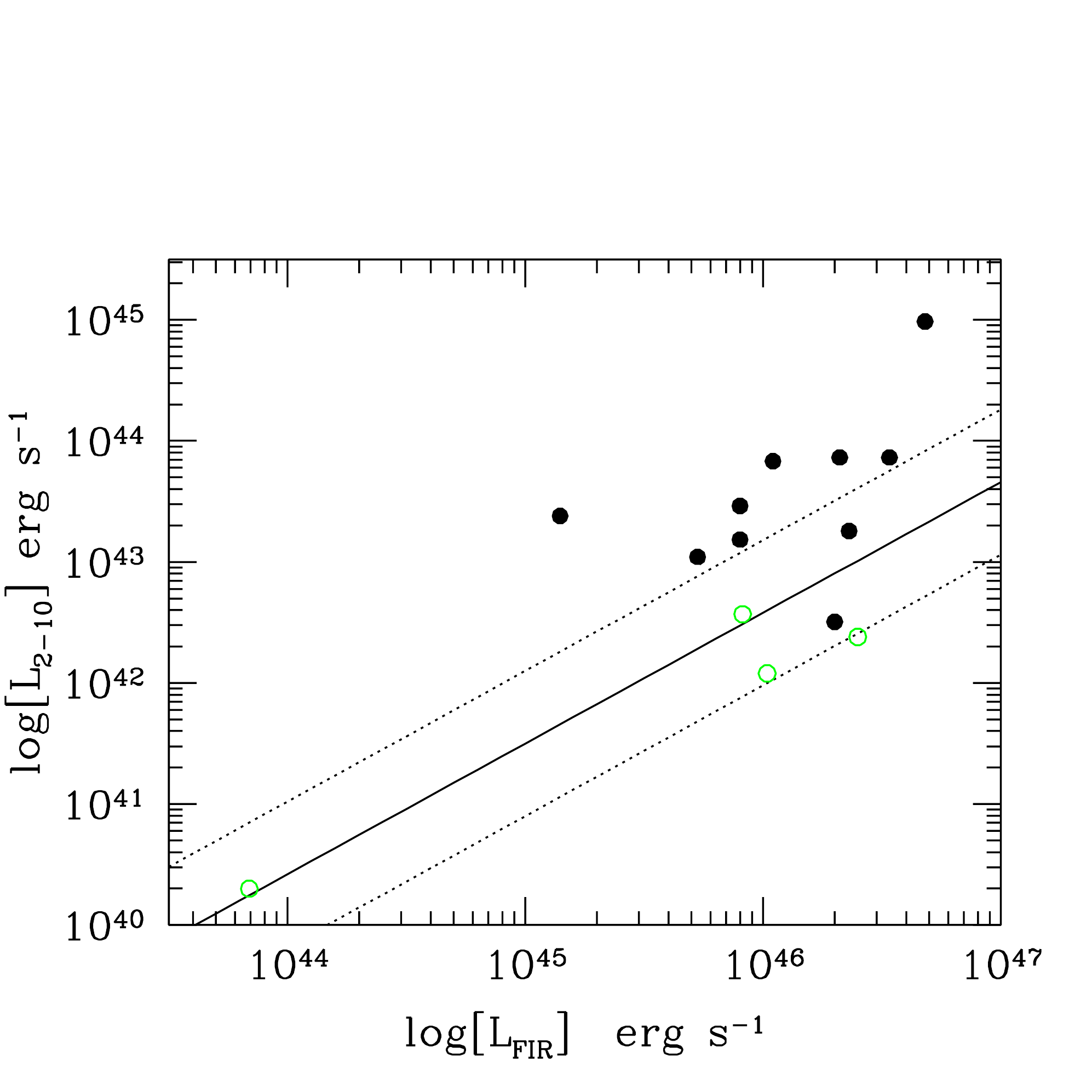} 
\caption{The X-ray (2-10 keV) vs. the FIR (4-100 $\rm \mu m$) relation for our sources. The solid and dotted lines denote 
 the best-fit relation of Ranalli et al. (2003) - and its associated $2\sigma$ error, for normal galaxies in the local Universe.
 Open circles denote the four sources classified as galaxies on the basis of the X-ray diagnostics.}
  \label{ranalli}
  \end{center}
\end{figure*}

\begin{figure*}
   \begin{center}
\includegraphics[width=9.cm]{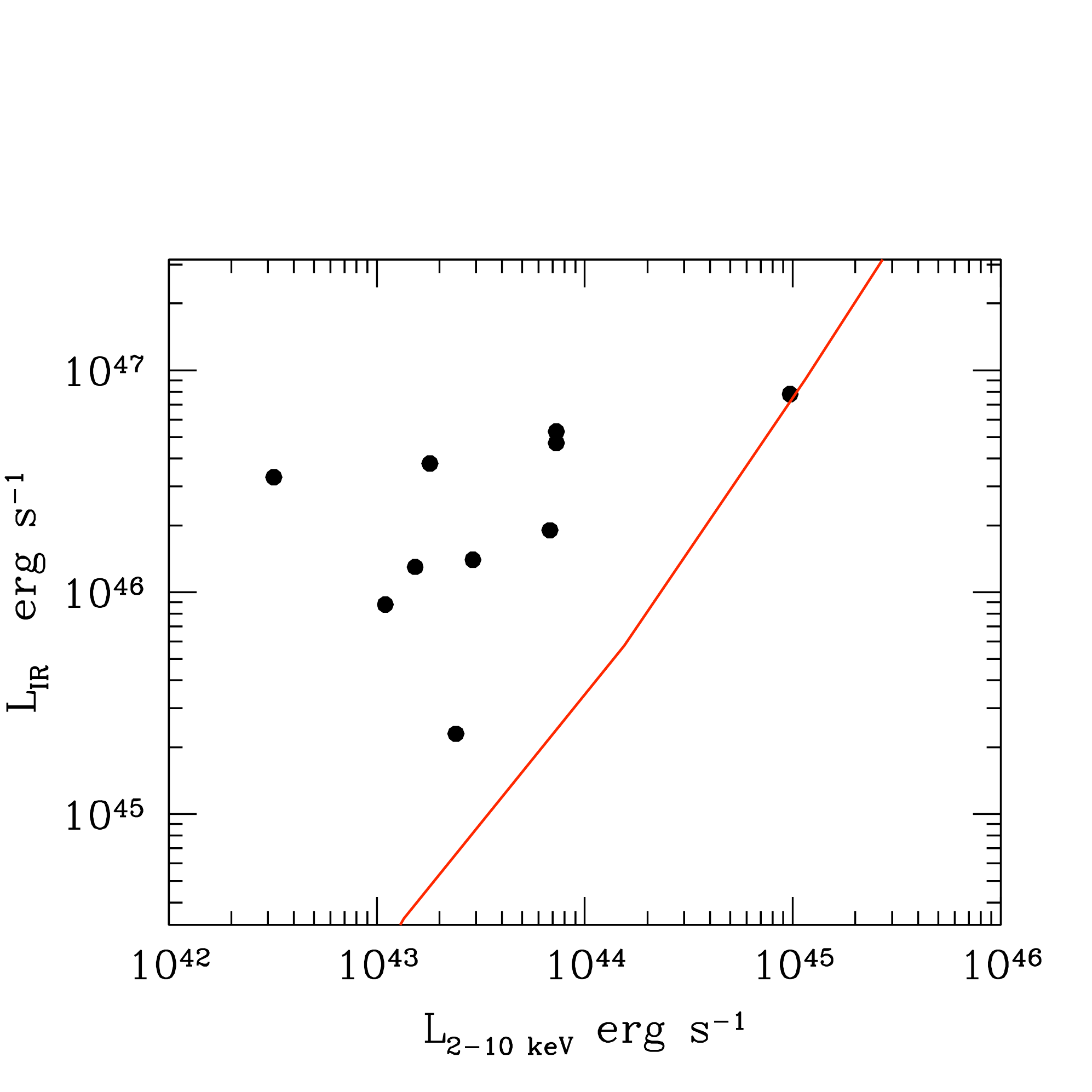} 
\caption{The X-ray (2-10 keV) vs. the total IR (10-1000 $\rm \mu m$) luminosity relation for our sources. The solid line denotes 
 the empirical relation of Hopkins et al. (2007) for the bolometric correction of AGN.}
  \label{hopkins}
  \end{center}
\end{figure*}

\section{Discussion} 
In the previous sections we presented a comprehensive analysis of the X-ray and mid-IR properties of the 
  SMG galaxies in the eCDFS and CDFS. These properties are used to 
   assess (1) the fraction of AGN among the SMG population  (2) the amount of 
     obscuration and in particular  Compton-thick obscuration 
       (3) The power-mechanism among the SMG which host an AGN 
     
     \subsection{Fraction of AGN among SMGs}
     \label{agn_fraction}
     Using robust positions from {\it Spitzer} MIPS, we find 
      14 associations of X-ray sources with LABOCA SMGs
       or a fraction of  ($11\pm 3$ \%).
        As the observations in the area of the CDFS 
         are very sensitive,  it is very likely that a significant number of sources 
          are associated with normal galaxies rather than AGN 
          (see Hornschemeier et al. 2003, 
           Tzanavaris \& Georgantopoulos 2008, Ptak et al. 2008 and references therein).    
     The X-ray analysis has suggested that a number (four) of 
      SMGs are associated with normal galaxies rather than AGN. 
       This was based on their relatively low X-ray luminosities 
       (less than a few times $10^{42}$\,\lunits) in combination 
        with unobscured X-ray spectra. 
        A further check on whether these sources are associated 
         with normal galaxies can be provided by the $L_{\rm x}-L_{\rm FIR}$ 
          relation of Ranalli et al. (2003).
           These authors have found a strong correlation between the 
  X-ray and the IR luminosity in star-forming galaxies in the local Universe. 
 This implies that only sources with an excess of X-ray emission 
  above this relation can be securely associated with AGN.
  We show this relation in Fig.\,\ref{ranalli}. 
   It can be seen that the four sources classified as galaxies  on the basis 
    of the X-ray diagnostics alone also follow the $L_{\rm x}-L_{\rm FIR}$ relation of Ranalli et al. (2003).
     A fifth source which falls  below the Ranalli et al. (2003) relation has a flat spectrum 
     and thus is most probably an AGN (W-114).

        Excluding the four candidate galaxies, 
         the fraction of X-ray detected AGN among the LABOCA SMGs amounts to
          10/126 ($8\pm 2)$\%. Constraining this analysis in the 
           CDFs where the sensitivity is the highest yields 
            a higher AGN fraction, as expected. In particular, there are 
             38 LABOCA sources in CDFS; out of these, ten are 
          associated with X-ray sources translating to a 
           percentage of X-ray detections of ($26\pm 9$)\%.  
          As among these ten sources, three are most likely associated with galaxies,    
      a conservative estimate for the AGN fraction amounts to 7/38 ($18\pm7$)\%.  
      
      We compare the figures above with the fraction of X-ray 
       sources and AGN in particular  
       that Alexander et al. (2005b) and Laird et al. (2010) 
        have derived in the CDFN. As the X-ray sensitivity 
         of the CDFS and CDFN are identical this comparison 
          is straightforward. Alexander et al. (2005b) find a very high 
           fraction of X-ray detections among radio-detected SMGs ($75\pm 19$\%). 
            Laird et al. (2010) find a lower fraction 
             of X-ray detections among the overall SMG population ($45\pm8$ \%). 
              The primary difference between the analysis of 
               Laird et al. (2010) and that of Alexander et al. (2005b) 
                is that the former uses a purely sub-mm selected sample
         (Pope et al. 2006) while the latter employs a mixture of SMGs 
         and radio sources with follow-up SCUBA photometry. Therefore the sample of
            Laird et al. (2010) which contains 35 SMGs 
            is more appropriate for comparison with our work.  
             Although this sample gives a higher fraction of X-ray detections 
              compared with our sample,   
            their AGN fraction of about 25$\pm8$ \% 
             is in reasonable agreement with the results presented here.  
      
      The AGN fraction derived here can be considered only as a lower limit
       to the true AGN fraction.  This is because there may be many more 
        low luminosity (for example obscured) AGN among the non-X-ray detected SMGs. 
         The X-ray stacking analysis could shed more light on such a possibility.  
         A strong signal is detected in the soft band in the eCDFS with  a
          corresponding flux of $2\times 10^{-17}$\,\funits. 
          If we adopt the average redshift of the non X-ray detections in the SMG sample
   of Pope et al. (2006) (z=2.1) as the median redshift of our non X-ray
  detected SMGs, we find that their X-ray luminosity amounts to
           $5\times 10^{41}$\,\lunits. 
           The luminosity obtained above is comparable to that obtained by Laird et al. (2010) 
           for their X-ray undetected SMGs in the soft band ($4\times 10^{41}$\,\lunits).  
          The hardness ratio of -0.40$\pm0.10$ shows relatively little absorption. In particular, 
           this hardness ratio corresponds 
          to $N_{\rm H}= 1\times10^{21}$\,\cunits at the observer's frame assuming a photon index 
           of $\Gamma=1.8$ (or $N_{\rm H}\approx 2\times10^{22}$\cunits at z=2). 
           Alternatively, this hardness ratio corresponds to  a photon-index of $\Gamma\approx 1.6$  (for $N_{\rm H}=0$).  
         Given that the implied spectrum  is relatively unobscured, it is reasonable to assume that the bulk of the X-ray emission  
          comes from star-forming processes and to estimate the star-formation-rate (SFR).  
                As the IR luminosity provides a robust measurement 
       of the star-formation rate (SFR), the tight $L_{\rm x}-L_{\rm FIR}$ correlation found 
        by Ranalli et al. (2003) implies that X-rays are 
        also a good tracer of the star-formation in galaxies. Based on their $L_X-L_{IR}-SFR$ 
       correlation, we find an average
        SFR of $\rm 110\,M_\odot\,yr^{-1}$  for our non-X-ray detected sources. 
    This is in agreement with the  value derived by Laird et al. (2010) namely 150$\rm M_\odot yr^{-1}$. 
     As these authors point out, this value is well below the huge SFR ($\sim 1000  \rm M_\odot yr^{-1}$ 
      which have been derived for SMGs on the basis of the FIR luminosities. 
      This is also below the SFR derived for the three most luminous candidate galaxies in our sample
       (W-45, W-59 and W-67) which ranges between $\sim$ 300-1000   $\rm M_\odot yr^{-1}$, 
        according to the X-ray luminosity and the  relation of Ranalli et al. (2003). 
        It is rather unlikely that the low SFR derived from the stacked signal in SMGs
         can be attributed to obscured star-formation.  
        For example, if the column density is $N_{\rm H}\sim10^{22}$\cunits at z=2
          as suggested above, the 0.5-2 keV luminosity and thus the SFR would increase by a factor of only 1.5.

     \subsection{Obscuration} 
     The direct X-ray spectral analysis reveals that the majority of AGN are 
     significantly absorbed having column densities above $10^{23}$ \cunits (see Fig.\,\ref{nh}).
     One of the  sources could be even classified as a transmission dominated Compton-thick AGN:
      this is W-11 at a photometric resdhift of z=6.07. Nevertheless, this classification is highly ambiguous
       because of the associated redshift uncertainties. The $L_{\rm x}-L_{\rm 6\,\mu m}$
        diagram suggests that source W-114 may be related to a reflection-dominated  Compton-thick source. 
     Its flat X-ray spectrum ($\Gamma < 1.32$ at the 90\% confidence limit) is also pointing towards 
      this interpretation. Our results on the amount of obscuration are in reasonable agreement 
       with those of Laird et al. (2010) in the CDFN. These authors find seven bona-fide AGN among their 18 
       SMG/X-ray associations. Their X-ray spectral analysis shows a median obscuration  
       of $\sim 10^{23}$\,\cunits with one of their sources being a Compton-thick AGN. 
     
     The stacking analysis provides significant constraints on the fraction of Compton-thick sources 
      among the SMG undetected in X-rays. The detected signal is relatively soft (hr=-0.40 between the 0.5-2\,keV and the 2-5\,keV bands),
       implying a spectrum with a photon index of $\Gamma=1.6$. 
       This immediately suggests a small fraction of hard spectrum sources such as those associated with Compton-thick AGN. 
        In order to quantify the fraction 
        of Compton-thick AGN,  we assume a simplistic model where the undetected SMG population consists 
         solely of star-forming galaxies and Compton-thick AGN. 
      We assume that normal galaxies have an average spectrum of $\Gamma=1.7$, (e.g. Zezas et al. 1998) 
      which corresponds to a hardness ratio of -0.44 in the above spectral bands. We further adopt that  
      Compton-thick sources have a spectrum of $\Gamma=1$ 
       (e.g. Matt et al. 2004, Georgantopoulos et al. 2009) or equivalently a hardness ratio of   -0.13.
      We derive a fraction of Compton-thick sources of about 10\%. 
       We point out  that this estimate is valid only under the assumption 
        that the X-ray fluxes from all the sources are roughly the same. 
      This fraction should only be considered as an upper limit. 
        In the case where there exist AGN with significant obscuration (but not necessarily Compton-thick)  among the undetected SMG sources,
         the fraction of Compton-thick sources will be lower. For example, AGN with a rest-frame column density 
          of $N_{\rm H}=10^{23}$\,\cunits at z=2 have a hardness ratio similar to that of a reflection-dominated Compton-thick AGN
           with $\Gamma=1$.

     \subsection{The SMG power-mechanism} 
     We discuss here the possibility that the AGN powers the 
     total IR emission in the case of our ten bona-fide AGN, i.e. we exclude the 
      four sources classified as possible galaxies on the basis 
       of low X-ray luminosities and soft X-ray spectra
        (see section\,\ref{agn_fraction} above).
         The SED fitting provides a clear picture of the 
          AGN contribution. We find that even in the two sources 
           which have power-law IRAC photometric distributions 
            and thus are clearcut AGN in the mid-IR part of the spectrum,
             the torus contributes less than 20\% of the 
              total IR (10-1000\,$\rm \mu m$) luminosity. 
           Laird et al. (2010) estimate  a fraction of 
                3/16 for AGN-powered SMGs in their sample.
                      
               In order to better visualize the level of the AGN contribution, 
               we plot the total IR luminosity as a function of the X-ray luminosity 
                in Fig.\,\ref{hopkins} for our sources together with the 
                 empirically estimated relation for AGN (Hopkins et al. 2007).
                 We see that only one source (W-9) follows this relation, while 
                  the remaining sources  are far above the relation of Hopkins et al. (2007), 
                   implying  a negligible AGN contribution.

     \section{Conclusions}
     We explore the X-ray properties of the 126 sub-mm sources of the LABOCA survey 
      in the area of the CDFS and eCDFS. Using {\it Spitzer} and radio counterparts 
      of the sub-mm sources where 
       available, we find 14 sources with significant X-ray detection. Our results can be 
        summarised as follows: 
        
        \begin{itemize}
        \item{The X-ray luminosities and spectra suggest that most of the sub-mm - X-rays associations 
         (10/14) host an AGN. In four sources the X-ray emission could 
          possibly originate from star-forming processes.}
          \item{The fraction of X-ray AGN among the LABOCA SMG sample in the area 
           of the CDF-S is at most  $26\pm9$ \%.}
          \item{Six of our X-ray sources show significant amounts of absorption $N_{\rm H}>10^{23}$ \cunits  
           but there is no unambiguous evidence  that any of the sources is a Compton-thick AGN.}
           \item{Detailed SED fitting shows that most of the AGN require a torus component.
           However, the AGN contribution to the total IR luminosity is small.}
           \item{X-ray stacking analysis  of the undetected SMGs reveals a signal 
            with a relatively soft spectrum. This is more suggestive of a SFR 
             galaxy population.} 
              \end{itemize}  
     
\begin{acknowledgements}  
IG and AC acknowledge the Marie Curie fellowship
 FP7-PEOPLE-IEF-2008 Prop. 235285.
 AC acknowledges receipt of ASI grants I/023/05/00 and I/88/06. 
We acknowledge the use of {\it Spitzer} data provided by the 
 {\it Spitzer} Science Center. 
 The Chandra data were taken from the Chandra Data Archive 
 at the Chandra X-ray Center.   
\end{acknowledgements}

\end{document}